\documentclass[aos,preprint]{imsart}
\pdfoutput=1

\usepackage[OT1]{fontenc}
\usepackage{natbib}
\usepackage{amsmath}
\usepackage{amssymb}
\usepackage{amsthm}
\usepackage{graphicx}
\usepackage{color}
\usepackage{subfig}
\usepackage{algorithm}
\usepackage[noend]{algpseudocode}
\usepackage{microtype}
\usepackage{url}

\startlocaldefs
\usepackage{macros}
\endlocaldefs

\setattribute{journal}{name}{}

\begin{document}
\begin{frontmatter}
  \title{Tree-Structured Stick Breaking Processes for Hierarchical Data}%
  \runtitle{Tree-Structured Stick Breaking Processes}%

  \begin{aug}

    \author{\fnms{Ryan P.} \snm{Adams}%
      \ead[label=e1]{rpa@cs.toronto.edu}},%
    \address{Department of Computer Science\\
      University of Toronto\\
      \printead{e1}}
    \author{\fnms{Zoubin} \snm{Ghahramani}%
      \ead[label=e2]{zoubin@eng.cam.ac.uk}}%
    \address{Department of Engineering\\
      University of Cambridge\\
      \printead{e2}}
    \and
    \author{\fnms{Michael I.} \snm{Jordan}%
      \ead[label=e3]{jordan@cs.berkeley.edu}}%
    \address{Department of Statistics\\
      University of California at Berkeley\\
      \printead{e3}}%

    \runauthor{R.P. Adams et al.}
  \end{aug}
  
  \begin{abstract}
    Many data are naturally modeled by an unobserved hierarchical
    structure.  In this paper we propose a flexible nonparametric prior
    over unknown data hierarchies.  The approach uses nested
    stick-breaking processes to allow for trees of unbounded width and
    depth, where data can live at any node and are infinitely
    exchangeable.  One can view our model as providing infinite mixtures
    where the components have a dependency structure corresponding to an
    evolutionary diffusion down a tree.  By using a stick-breaking
    approach, we can apply Markov chain Monte Carlo methods based on
    slice sampling to perform Bayesian inference and simulate from the
    posterior distribution on trees.  We apply our method to
    hierarchical clustering of images and topic modeling of text data.
  \end{abstract}
\end{frontmatter}

\section{Introduction}
Structural aspects of models are often critical to obtaining flexible,
expressive model families.  In many cases, however, the structure is 
unobserved and must be inferred, either as an end in itself or to assist 
in other estimation and prediction tasks.  This paper addresses an important 
instance of the structure learning problem: the case when the data arise
from a latent hierarchy.  We take a direct nonparametric Bayesian approach, 
constructing a prior on tree-structured partitions of data that provides 
for unbounded width and depth while still allowing tractable posterior 
inference.

Probabilistic approaches to latent hierarchies have been explored in
a variety of domains.  Unsupervised learning of densities and
nested mixtures has received particular attention via finite-depth
trees \citep{williams-2000a}, diffusive branching processes
\citep{neal-2003a} and hierarchical clustering
\citep{heller-ghahramani-2005a,teh-etal-2007a}.  Bayesian approaches to
learning latent hierarchies have also been useful for semi-supervised
learning \citep{kemp-etal-2004a}, relational learning
\citep{roy-etal-2007a} and multi-task learning \citep{daume-2009a}.  In
the vision community, distributions over trees have been useful as
priors for figure motion \citep{meeds-etal-2008a} and for discovering
visual taxonomies \citep{bart-etal-2008a}.

In this paper we develop a distribution over probability measures that
imbues them with a natural hierarchy.  These hierarchies have
unbounded width and depth and the data may live at internal nodes on
the tree.  As the process is defined in terms of a distribution over
probability measures and not as a distribution over data per se, data
from this model are infinitely exchangeable; the probability of any
set of data is not dependent on its ordering.  Unlike other infinitely
exchangeable models \citep{neal-2003a,teh-etal-2007a}, a pseudo-time
process is not required to describe the distribution on trees and it
can be understood in terms of other popular Bayesian nonparametric
models.

Our new approach allows the components of an infinite mixture model to
be interpreted as part of a diffusive evolutionary process.  Such a
process captures the natural structure of many data.  For example,
some scientific papers are considered \textit{seminal} --- they spawn
new areas of research and cause new papers to be written.  We might
expect that within a text corpus of scientific documents, such papers
would be the natural ancestors of more specialized papers that
followed on from the new ideas.  This motivates two desirable features
of a distribution over hierarchies: 1)~ancestor data (the
``prototypes'') should be able to live at internal nodes in the tree,
and 2)~as the ancestor/descendant relationships are not known \textit{a
  priori}, the data should be infinitely exchangeable.

\section{A Tree-Structured Stick-Breaking Process}
\label{sec:tssb}
\begin{figure}[t!]
  \centering%
  \subfloat[Dirichlet process stick breaking]{%
    \label{fig:stick-schematic:dp}%
    \includegraphics[width=\textwidth]{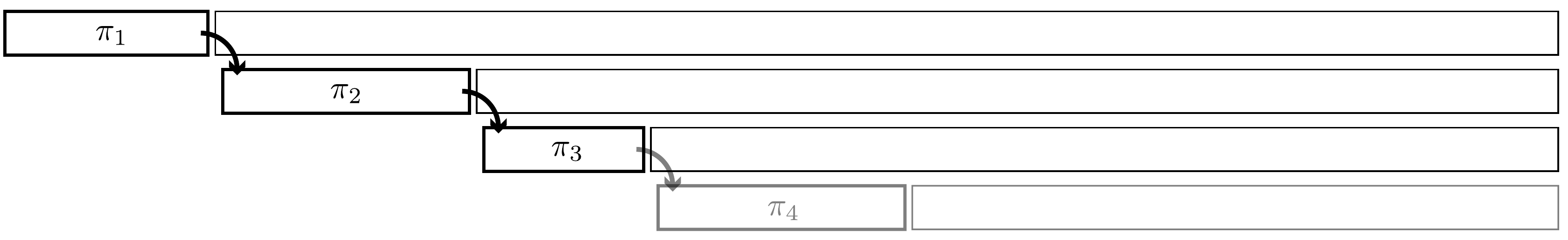}%
  }\\%
  \subfloat[Tree-structured stick breaking]{%
    \label{fig:stick-schematic:tssb}%
    \includegraphics[width=\textwidth]{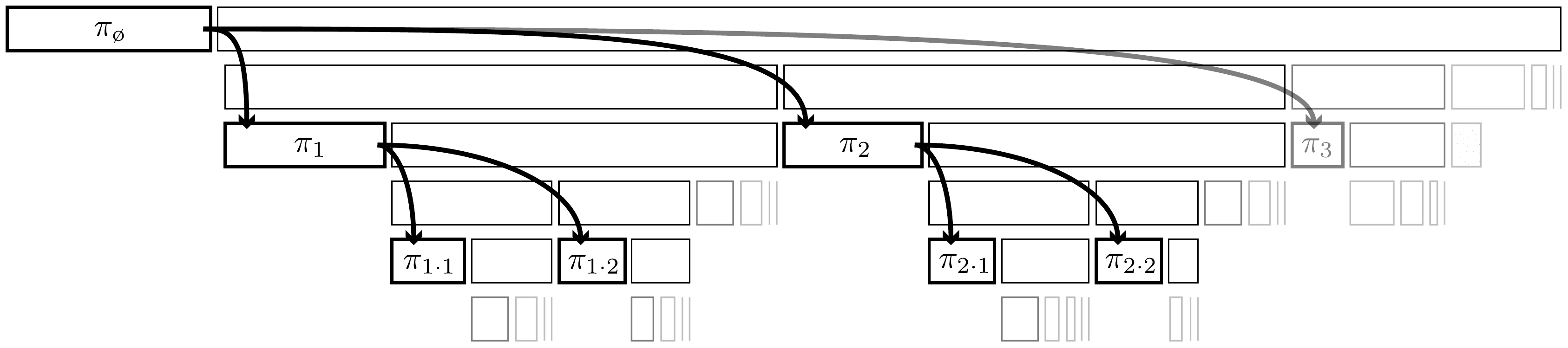}%
  }%
  \caption{a)~Dirichlet process stick-breaking procedure, with
    a linear partitioning.  b)~Interleaving two stick-breaking
    processes yields a tree-structured partition.  Rows 1, 3 and 5
    are $\nu$-breaks.  Rows 2 and 4 are $\psi$-breaks.}
  \label{fig:stick-schematic}
\end{figure}

Stick-breaking processes based on the beta distribution have played a
prominent role in the development of Bayesian nonparametric methods, most
significantly with the constructive approach to the Dirichlet process (DP)
due to \citet{sethuraman-1994a}.  A random probability measure~$G$
can be drawn from a DP with base measure~$\alpha H$ using a sequence of
beta variates via:
\begin{align}
  \label{eqn:dp}
  G &= \sum^{\infty}_{i=1}\pi_i\,\delta_{\theta_i} &
  \pi_i &= \nu_i\prod^{i-1}_{i'=1}(1-\nu_{i'})
\end{align}
\vskip -0.4cm%
\begin{align*}
  \theta_i &\sim H &
  \nu_i &\sim \distBeta(1,\alpha) &
  \pi_1&=\nu_1.
\end{align*}
We can view this as taking a stick of unit length and breaking it at a
random location.  We call the left side of the stick~$\pi_1$ and then
break the right side again at a new place, calling the left side of
this new break~$\pi_2$.  If we continue this process of ``keep the
left piece and break the right piece again'' as in
Fig.~\ref{fig:stick-schematic:dp}, assigning each~$\pi_i$ a random
value drawn from~$H$, we can view this is a random probability measure
centered on~$H$.  The distribution over the
sequence~${(\pi_1,\pi_2,\cdots)}$ is a case of the GEM distribution
\citep{pitman-2002a}, which also includes the Pitman-Yor process
\citep{pitman-yor-1997a}.  Note that in Eq.~(\ref{eqn:dp}) the~$\theta_i$
are i.i.d.\ from~$H$; in the current paper these parameters will be
drawn according to a hierarchical process.

The GEM construction provides a distribution over infinite partitions of
the unit interval, with natural numbers as the index set as in
Fig.~\ref{fig:stick-schematic:dp}.  In this paper, we extend this idea to
create a distribution over infinite partitions that also possess a
hierarchical graph topology.  To do this, we will use finite-length
sequences of natural numbers as our index set on the partitions.  Borrowing
notation from the P{\'o}lya tree (PT) construction
\citep{mauldin-etal-1992a}, let~${\bepsilon \seq (\epsilon_1,
  \epsilon_2,\cdots,\epsilon_K)}$, denote a length-$K$ sequence of positive
integers, i.e.,~${\epsilon_k\shortin\naturals^{+}}$.  We denote the
zero-length string as~${\bepsilon\seq\emptystr}$ and use~$|\bepsilon|$ to
indicate the length of~$\bepsilon$'s sequence.  These strings will index
the nodes in the tree and~$|\bepsilon|$ will then be the depth of
node~$\bepsilon$.

We interleave two stick-breaking procedures as in
Fig.~\ref{fig:stick-schematic:tssb}.  The first has beta variates
${\nu_{\bepsilon}\ssim\distBeta(1,\alpha(|\bepsilon|))}$ which determine
the size of a given node's partition as a function of depth.  The second
has beta variates~${\psi_{\bepsilon}\ssim\distBeta(1,\gamma)}$, which
determine the branching probabilities.  Interleaving these processes
partitions the unit interval.  The size of the partition associated with
each~$\bepsilon$ is given by
\begin{align}
  \label{eqn:tssb}%
  \pi_{\bepsilon} &= \nu_{\bepsilon}\varphi_{\bepsilon}
  \prod_{\bepsilon'\prec\bepsilon}\varphi_{\bepsilon'}(1-\nu_{\bepsilon'})
  &
  \varphi_{\bepsilon\epsilon_i} &= \psi_i\prod^{i-1}_{i'=1}(1-\psi_{i'})
  &
  \pi_{\emptystr} &=\nu_{\emptystr},
\end{align}
where~$\bepsilon\epsilon_i$ denotes the sequence that results from
appending~$\epsilon_i$ onto the end of~$\bepsilon$,
and~${\bepsilon'\sprec\bepsilon}$ indicates that~$\bepsilon$ could be
constructed by appending onto~$\bepsilon'$.  When viewing these strings as
identifying nodes on a tree,~${\{\bepsilon\epsilon_i : \epsilon_i \shortin
  1,2,\cdots\}}$ are the children of~$\bepsilon$ and~${\{\bepsilon':
  \bepsilon'\sprec\bepsilon\}}$ are the ancestors of~$\bepsilon$.
The~$\{\pi_{\bepsilon}\}$ in Eq.~(\ref{eqn:tssb}) can be seen as products of
several decisions on how to allocate mass to nodes and branches in the tree:
the~$\{\varphi_{\bepsilon}\}$ determine the probability of a particular
sequence of children and the~$\nu_{\bepsilon}$ and~$(1\sminus\nu_{\bepsilon})$
terms determine the proportion of mass allotted to~$\bepsilon$ versus nodes
that are descendants of~$\bepsilon$.

We require that the~$\{\pi_{\bepsilon}\}$ sum to one.  The~$\psi$-sticks
have no effect upon this, but~${\alpha(\cdot):\naturals\to\reals^{+}}$ (the
depth-varying parameter for the $\nu$-sticks) must satisfy
${\sum_{j\seq1}^{\infty}\ln(1 \splus 1/\alpha(j\sminus1))\seq+\infty}$ (see
\citet{ishwaran-james-2001a}).  This is clearly true
for~$\alpha(j)\seq\alpha_0\sgt 0$.  A useful function that also satisfies
this condition is~$\alpha(j)\seq\lambda^{j}\alpha_0$ with~$\alpha_0\sgt 0,
\lambda\shortin(0,1]$.  The decay parameter~$\lambda$ allows a distribution
  over trees with most of the mass at an intermediate depth.  This is
  the~$\alpha(\cdot)$ we will assume throughout the remainder of the paper.

\subsubsection*{An Urn-based View}
When a Bayesian nonparametric model induces partitions over data, it
is sometimes possible to construct an urn scheme that corresponds to
sequentially generating data, while integrating out the underlying
random measure.  The ``Chinese restaurant'' metaphor for the Dirichlet
process is a popular example.  In our model, we can use such an urn
scheme to construct a treed partition over a finite set of data.  Note
that while the tree illustrated in
Fig.~\ref{fig:stick-schematic:tssb} is a nested set of size-biased
partitions, the ordering of the branches in an urn-based tree over
data does not necessarily correspond to a size-biased permutation
\citep{pitman-1996a}. 

\begin{figure}[t!]
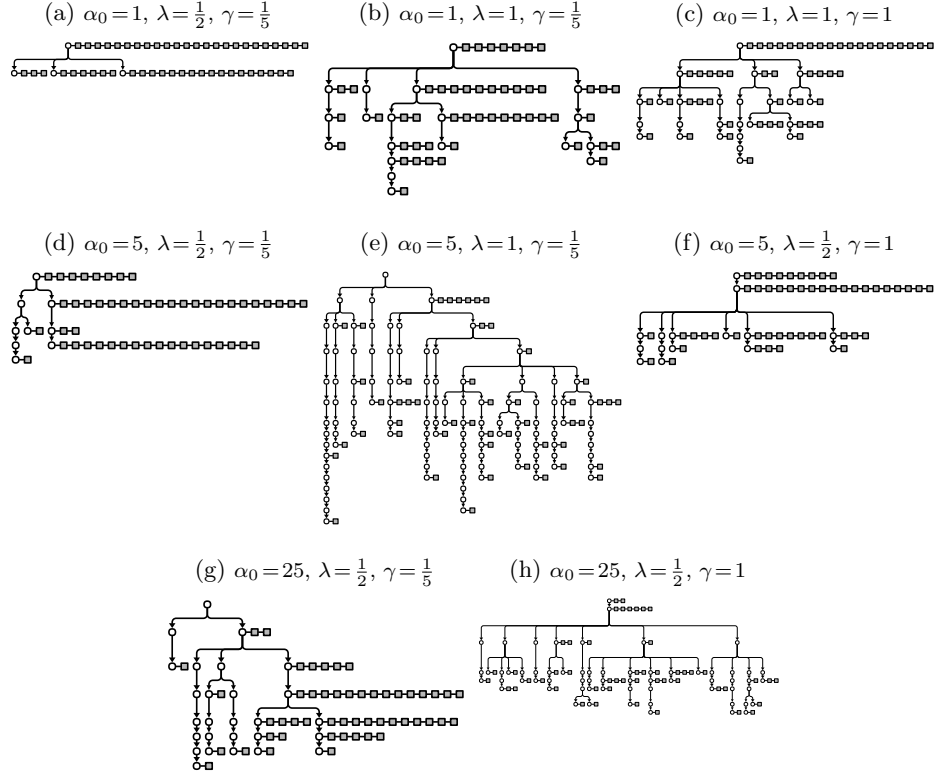

  \centering%
  \captionsetup[subfloat]{position=top}
  \newcommand{\priordraw}[4]{figures/prior-draws/prior-draw-alpha#1-gamma#2-decay#3-seed#4.pdf}
  \subfloat[{\scriptsize $\alpha_0\seq 1$, $\lambda\seq\half$, $\gamma\seq\fifth$}]{%
    \includegraphics[width=0.32\textwidth]{\priordraw{1.00}{0.20}{0.50}{1001}}%
  }~%
  \subfloat[{\scriptsize $\alpha_0\seq 1$, $\lambda\seq 1$, $\gamma\seq\fifth$}]{%
    \includegraphics[width=0.32\textwidth]{\priordraw{1.00}{0.20}{1.00}{1092}}%
  }~%
  \subfloat[{\scriptsize $\alpha_0\seq 1$, $\lambda\seq 1$, $\gamma\seq 1$}]{%
    \includegraphics[width=0.32\textwidth]{\priordraw{1.00}{1.00}{1.00}{1014}}%
  }\\%
  \subfloat[{\scriptsize $\alpha_0\seq 5$, $\lambda\seq\half$, $\gamma\seq\fifth$}]{%
    \includegraphics[width=0.32\textwidth]{\priordraw{5.00}{0.20}{0.50}{1001}}%
  }~%
  \subfloat[{\scriptsize $\alpha_0\seq 5$, $\lambda\seq 1$, $\gamma\seq\fifth$}]{%
    \includegraphics[width=0.32\textwidth]{\priordraw{5.00}{0.20}{1.00}{1012}}%
  }~%
  \subfloat[{\scriptsize $\alpha_0\seq 5$, $\lambda\seq \half$, $\gamma\seq 1$}]{%
    \includegraphics[width=0.32\textwidth]{\priordraw{5.00}{1.00}{0.50}{1003}}%
  }\\%
  \subfloat[{\scriptsize $\alpha_0\seq 25$, $\lambda\seq \half$, $\gamma\seq\fifth$}]{%
    \includegraphics[width=0.32\textwidth]{\priordraw{25.00}{0.20}{0.50}{1001}}%
  }~%
  \subfloat[{\scriptsize $\alpha_0\seq 25$, $\lambda\seq \half$, $\gamma\seq 1$}]{%
    \includegraphics[width=0.32\textwidth]{\priordraw{25.00}{1.00}{0.50}{1003}}%
  }%
  \label{fig:prior-draws}%
  \caption{Eight samples of trees over partitions of fifty
    data, with different hyperparameter settings.  The circles are
    represented nodes, and the squares are the data.  Note that some
    of the sampled trees have represented nodes with no data
    associated with them and that the branch ordering does not
    correspond to a size-biased permutation.}
\end{figure}

The data drawing process can be seen as a path-reinforcing Bernoulli trip
down the tree where each datum starts at the root and descends into
children until it stops at some node.  The first datum lives at the root
node with probability~${1/(\alpha(0) \splus 1)}$, otherwise it descends and
instantiates a new child.  It stays at this new child with
probability~${1/(\alpha(1)\splus 1)}$ or descends again and so on.  A later
datum stays at node~$\bepsilon$ with probability~${(N_{\bepsilon} \splus
  1)/(N_{\bepsilon} \splus N_{\bepsilon\prec\cdot} \splus
  \alpha(|\bepsilon|) \splus 1)}$, where~$N_{\bepsilon}$ is the number of
previous data that stopped at~$\bepsilon$, and~$N_{\bepsilon\prec\cdot}$ is
the number of previous data that came down this path of the tree but did
not stop at~$\bepsilon$, i.e., a sum over all
descendants:~${N_{\bepsilon\prec\cdot}\seq
  \sum_{\bepsilon\prec\bepsilon'}N_{\bepsilon'}}$.  If a datum descends
to~$\bepsilon$ but does not stop then it chooses which child to descend
to according to a Chinese restaurant process where the previous customers
are only those data who have also descended to this point.  That is, if it
has reached node~$\bepsilon$ but will not stay there, it descends to
existing child~$\bepsilon\epsilon_i$ with
probability~${(N_{\bepsilon\epsilon_i} \splus
  N_{\bepsilon\epsilon_i\prec\cdot})/(N_{\bepsilon\prec\cdot}\splus\gamma)}$
and instantiates a new child with
probability~${\gamma/(N_{\bepsilon\prec\cdot}\splus\gamma)}$.  A particular
path therefore becomes more likely according to its ``popularity'' with
previous data.  Note that a node can be a part of a popular path without
having any data of its own.  Fig.~\ref{fig:prior-draws} shows the
structures implied over fifty data drawn from this process with different
hyperparameter settings.

The urn view allows us to place this model into the literature on priors on
infinite trees.  One of the main contributions of this model is that the
data can live at arbitrary internal nodes in the tree, but are nevertheless
infinitely exchangeable.  This is in contrast to the model proposed by
\citet{meeds-etal-2008a}, for example, which is not infinitely exchangeable.
The nested Chinese restaurant process~(nCRP)~\citep{blei-etal-2010a}
provides a distribution over trees of unbounded width and depth, but data
correspond to reinforcing paths of infinite length, requiring an additional
distribution over depths that is not path-dependent.  The P{\'o}lya tree
\citep{mauldin-etal-1992a} uses a recursive stick-breaking process to
specify a distribution over nested partitions in a binary tree, however the
resulting data live at the infinitely-deep leaf nodes.  The marginal
distribution over the topology of a Dirichlet diffusion tree
\citep{neal-2003a} (and the clustering variant of Kingman's coalescent proposed by
\citet{teh-etal-2007a}) provides path-reinforcement and infinite
exchangeability, however the topology is determined by a hazard process in
pseudo-time and data do not live at internal nodes.

\section{Hierarchical Priors for Node Parameters}
\label{sec:node-params}
In the stick-breaking construction of the Dirichlet process one can view
the procedure as generating an infinite partition and then labeling each
cell~$i$ with parameter~$\theta_i$ drawn i.i.d.\ from~$H$.  In a mixture
model, data that are drawn from the~$i$th component are generated
independently according to a distribution~$f(x\given\theta_i)$, where~$x$
takes values in a sample space~$\mcX$.  In our model, we continue to assume
that the data are generated independently given the latent labeling, but to
take advantage of the tree-structured partitioning of
Section~\ref{sec:tssb} an i.i.d.\ assumption on the node parameters is
inappropriate.  Rather, the distribution over the parameters at
node~$\bepsilon$, denoted~$\theta_{\bepsilon}$, should
depend in an interesting way on its
ancestors~${\{\theta_{\bepsilon'}:\bepsilon'\sprec\bepsilon\}}$.  A natural
and powerful way to specify such dependency is via a directed graphical
model, with the requirement that edges must always point down the tree.  An
intuitive subclass of such graphical models are those in which a child is
conditionally independent of all ancestors, given its parents and any
global hyperparameters.  This is the case we will focus on here, as it
provides a useful view of the parameter-generation process as a ``diffusion
down the tree'' via a Markov transition kernel that can be essentially any
distribution with a location parameter.  Coupling such a kernel, which we
denote~$T(\theta_{\bepsilon\epsilon_i}\sgets\theta_{\bepsilon})$, with a
root-level prior~$p(\theta_{\emptystr})$ and the node-wise data
distribution~$f(x\given\theta_{\bepsilon})$, we have a complete model for
infinitely exchangeable tree-structured data on~$\mcX$.  We now examine a
few specific examples.

\paragraph*{Generalized Gaussian Diffusions}
If our data distribution~$f(x\given\theta)$ is such that the
parameters can be specified as a real-valued
vector~$\theta\shortin\reals^M$, then we can use a Gaussian distribution to
describe the parent-to-child transition
kernel:~${T_{\textsf{norm}}(\theta_{\bepsilon\epsilon_i}\sgets\theta_{\bepsilon})
  \seq
  \distNorm(\theta_{\bepsilon\epsilon_i}\given\eta\,\theta_{\bepsilon},
  \Lambda)}$, where~${\eta\shortin[0,1)}$.  Such a kernel captures the
simple idea that the child's parameters are noisy versions of the
parent's, as specified by the covariance matrix~$\Lambda$,
while~$\eta$ ensures that all parameters in the tree have a finite
marginal variance.  While this will not result in a conjugate model
unless the data are themselves Gaussian, it has the simple property
that each node's parameter has a Gaussian prior that is specified by
its parent.  We present an application of this model in Section~\ref{sec:cifar},
where we model images as a distribution over binary vectors obtained by 
transforming a real-valued vector to~$(0,1)$ via the logistic function.

\paragraph*{Chained Dirichlet-Multinomial Distributions}
If each datum is a set of counts over~$M$ discrete outcomes, as in
many finite topic models, a multinomial model for~$f(x\given\theta)$
may be appropriate.  In this case,~${\mcX\seq \naturals^M}$,
and~$\theta_{\bepsilon}$ takes values in the~$(M\sminus 1)$-simplex.
We can construct a parent-to-child transition kernel via a Dirichlet
distribution with concentration parameter~$\kappa$:
${T_{\textsf{dir}}(\theta_{\bepsilon\epsilon_i}\sgets\theta_{\bepsilon})\seq\distDir(\kappa\theta_{\bepsilon})}$,
using a symmetric Dirichlet for the root node,
i.e.,~${\theta_{\emptystr}\ssim\distDir(\kappa\boldsymbol{1})}$.

\paragraph*{Hierarchical Dirichlet Processes}
A very general way to specify the distribution over data is to say
that it is a random probability measure drawn from a Dirichlet
process.  In our case, a very flexible model would say that the data
drawn at node~$\bepsilon$ are from a distribution~$G_{\bepsilon}$ as
in Eq.~(\ref{eqn:dp}).  This means that~$\theta_{\bepsilon}\sim
G_{\bepsilon}$ where~$\theta_{\bepsilon}$ now corresponds to an
infinite set of parameters.  The hierarchical Dirichlet process (HDP)
\citep{teh-etal-2006a} provides a natural parent-to-child transition
kernel for the tree-structured model, again with concentration
parameter~$\kappa$:~${T_{\textsf{hdp}}(G_{\bepsilon\epsilon_i}\sgets
  G_{\bepsilon}) \seq \distDP(\kappa G_{\bepsilon})}$.  At the top
level, we specify a global base measure~$H$ for the root node,
i.e.,~${G_{\emptystr}\ssim H}$.  One negative aspect of this
transition kernel is that the~$G_{\bepsilon}$ will have a tendency to
collapse down onto a single atom.  One remedy is to smooth the
kernel with~$\eta$ as in the Gaussian case,
i.e., ${T_{\textsf{hdp}}(G_{\bepsilon\epsilon_i}\sgets G_{\bepsilon})
  \seq \distDP(\kappa\, (\eta\, G_{\bepsilon} +(1-\eta)\,H))}$.

\section{Inference via Markov chain Monte Carlo}
\label{sec:mcmc}
We have so far defined a model for data that are generated from the
parameters associated with the nodes of a random tree.  Having seen~$N$
data points and assuming a model~$f(x\given\theta_{\bepsilon})$ as in the previous
section, we wish to infer possible trees and model parameters.  As in most
complex probabilistic models, closed form inference is impossible and we
instead perform inference by generating posterior samples via Markov chain
Monte Carlo (MCMC).  To operate efficiently over a variety of regimes
without tuning, we use slice sampling \citep{neal-2003b} extensively.
This allows us to sample from the true posterior distribution over the 
finite quantities of interest despite the fact that our  our model 
technically contains an infinite number of parameters.
The primary data structure in our Markov chain is
the set of~$N$ strings describing the current assignments of data to nodes,
which we denote~$\{\bepsilon_n\}^N_{n=1}$.  We represent the $\nu$-sticks
and parameters~$\theta_{\bepsilon}$ for all nodes that are traversed by the
data in its current assignments,
i.e., ${\{\nu_{\bepsilon},\theta_{\bepsilon}:\exists n,
  \bepsilon\sprec\bepsilon_n\}}$.  We additionally represent all
$\psi$-sticks in the ``hull'' of the tree that contains the data: if at
some node~$\bepsilon$ one of the~$N$ data paths passes through
child~$\bepsilon\epsilon_i$, then we represent all the $\psi$-sticks in the
set~${\bigcup_{\bepsilon_n}\bigcup_{\bepsilon\epsilon_i\preceq\bepsilon_n}
  \{\psi_{\bepsilon\epsilon_j}:\epsilon_j\sleq\epsilon_i\}}$.  We also
sample from the hyperparameters~$\alpha_0$,~$\gamma$, and~$\lambda$ for the
tree and any parameters associated with the likelihoods.

\begin{figure*}[t!]
  \def\algorithmicindent{0.2cm}
  \begin{minipage}[t]{0.49\textwidth}
    \begin{algorithmic}
      \Function{samp-assignment}{$n$}
      \State $p_{\sf{slice}} \sim \distUni(0,
      f(x_n\given\theta_{\bepsilon_n}))$
      \State $u_{\sf{min}}\gets 0$, $u_{\sf{max}}\gets 1$
      \Loop
      \State $u \sim \distUni(u_{\sf{min}},u_{\sf{max}})$
      \State $\bepsilon\gets$ \Call{find-node}{$u$, $\emptystr$}
      \State $p \gets f(x_n\given\theta_{\bepsilon})$
      \If{$p > p_{\sf{slice}}$}
      \Return $\bepsilon$
      \ElsIf{$\bepsilon \slt \bepsilon_n$} $u_{\sf{min}}\sgets u$
      \Else $\;\;u_{\sf{max}} \gets u$
      \EndIf
      \EndLoop
      \EndFunction
    \end{algorithmic}
  \end{minipage}~%
  \begin{minipage}[t]{0.49\textwidth}
    \begin{algorithmic}
      \Function{find-node}{$u$, $\bepsilon$}
      \If{$u < \nu_{\bepsilon}$}
      \Return $\bepsilon$
      \Else
      \State $u\gets (u-\nu_{\bepsilon})/(1-\nu_{\bepsilon})$
      \While{$u \slt 1\sminus \prod_{j}(1\sminus\psi_{\bepsilon\epsilon_j})$}
      \State Draw a new $\psi$-stick
      \EndWhile
      \State $\boldsymbol{e}\sgets$ edges from~$\psi$-sticks
      \State $i\sgets$ bin index for~$u$ from edges
      \State Draw~$\theta_{\bepsilon\epsilon_i}$ and~$\nu_{\bepsilon\epsilon_i}$ if necessary
      \State $u\gets (u-e_i)/(e_{i+1}-e_i)$
      \State \Return \Call{find-node}{$u$, $\bepsilon\epsilon_i$}
      \EndIf
      \EndFunction
    \end{algorithmic}
  \end{minipage}\\%
  \begin{minipage}{0.49\textwidth}
    \vskip 0.25cm%
    \begin{algorithmic}      
      \Function{size-biased-perm}{$\bepsilon$}
      \State $\rho\gets\emptyset$
      \While{represented children}
      \State $\boldsymbol{w}\gets$ weights from~$\{\psi_{\bepsilon\epsilon_i}\}$
      \State $\boldsymbol{w}\gets \boldsymbol{w}\backslash\rho$
      \State $j \sim \boldsymbol{w}$
      \State $\rho\gets$ append~$j$
      \EndWhile
      \State \Return $\rho$
      \EndFunction
    \end{algorithmic}
  \end{minipage}
  \vskip 0.25cm%
  \hrule
\end{figure*}

\paragraph{Slice Sampling Data Assignments}
The primary challenge in inference with Bayesian nonparametric mixture
models is often sampling from the posterior distribution over assignments,
as it is frequently difficult to integrate over the infinity of
unrepresented components.  To avoid this difficulty, we use a slice
sampling approach that can be viewed as a combination of the Dirichlet
slice sampler of \citet{walker-2007a} and the retrospective sampler
of \citet{papaspiliopoulos-roberts-2008a}.

Section~\ref{sec:tssb} described a path-reinforcing process for
generating data from the model.  An alternative method is to
draw a uniform variate~$u$ on~$(0,1)$ and break sticks until we know
what~$\pi_{\bepsilon}$ the $u$ fell into.  One can imagine throwing a
dart at the top of Fig.~\ref{fig:stick-schematic:tssb} and considering
which~$\pi_{\bepsilon}$ it hits.  We would draw the sticks and
parameters from the prior, as needed, conditioning on the state
instantiated from any previous draws and with parent-to-child
transitions enforcing the prior downwards in the tree.  Calling the
pseudocode function \textsc{find-node}($u$,~$\bepsilon$) with
${u\ssim\distUni(0,1)}$ and ${\bepsilon\seq\emptystr}$ draws such a
sample.  This provides a retrospective slice sampling scheme on~$u$,
allowing us to draw posterior samples without having to specify any
tuning parameters.

To slice sample the assignment of the $n$th datum, currently assigned
to~$\bepsilon_n$, we initialize our slice sampling bounds to~$(0,1)$.
We draw a new~$u$ from the bounds and use the \textsc{find-node}
function to determine the associated~$\bepsilon$ from the
currently-represented state, plus any additional state that must be
drawn from the prior.  We do a \textit{lexical} comparison
(``string-like'') of the new~$\bepsilon$ and our current
state~$\bepsilon_n$, to determine whether this new path corresponds to
a~$u$ that is ``above'' or ``below'' our current state.  This lexical
comparison prevents us from having to represent the initial~$u_n$.  We
shrink the slice sampling bounds appropriately, depending on the
result of the comparison, until we find a~$u$ whose assignment
satisfies the slice.  This procedure is given in pseudocode as
\textsc{samp-assignment}($n$).  After performing this procedure, we
can discard any state that is not in the previously-mentioned hull of
representation.

\paragraph{Gibbs Sampling Stick Lengths} Given the represented sticks and
the current assignments of nodes to data, it is straightforward to resample
the lengths of the sticks from the posterior beta distributions
\begin{align*}
  \nu_{\bepsilon}\given\textrm{data} 
  &\sim\distBeta(N_{\bepsilon} \splus 1, N_{\bepsilon\prec\cdot} \splus
  \alpha(|\bepsilon|))
  \\
  \psi_{\bepsilon\epsilon_i}\given\textrm{data} 
  &\sim \distBeta(N_{\bepsilon\epsilon_i\prec\cdot}\splus 1,
 \gamma\splus\textstyle{\sum_{j>i}N_{\bepsilon\epsilon_j\prec\cdot}}),
\end{align*}
where~$N_{\bepsilon}$ and~$N_{\bepsilon\prec\cdot}$ are the path-based
counts as described in Section~\ref{sec:tssb}.

\paragraph{Gibbs Sampling the Ordering of the $\psi$-Sticks} When
using the stick-breaking representation of the Dirichlet process, it is
crucial for mixing to sample over possible orderings of the sticks.  In our
model, we include such moves on the~$\psi$-sticks.  We iterate over each
instantiated node~$\bepsilon$ and perform a Gibbs update of the ordering of
its immediate children using its invariance under size-biased permutation
(SBP) \citep{pitman-1996a}.  For a given node, the~$\psi$-sticks provide a
``local'' set of weights that sum to one. We repeatedly draw without
replacement from the discrete distribution implied by the weights and keep
the ordering that results.  Pitman \citep{pitman-1996a} showed that
distributions over sequences such as our~$\psi$-sticks are invariant under
such permutations and we can view the
\textsc{size-biased-perm}($\bepsilon$) procedure as a Metropolis--Hastings
proposal with an acceptance ratio that is always one.

\paragraph{Slice Sampling Stick-Breaking Hyperparameters}
Given all of the instantiated sticks, we slice sample from the conditional
posterior distribution over the hyperparameters~$\alpha_0$,~$\lambda$
and~$\gamma$:
\begin{align*}
  p(\alpha_0,\lambda\given\{\nu_{\bepsilon}\}) &\propto 
  \indic(\alpha^{\sf{min}}_0\slt\alpha_0\slt\alpha^{\sf{max}}_0)
  \indic(\lambda^{\sf{min}}\slt\lambda\slt\lambda^{\sf{max}})
  \prod_{\bepsilon}\distBeta(\nu_{\bepsilon}\given 1,
  \lambda^{|\bepsilon|}\alpha_0)
  \\
  p(\gamma\given\{\psi_{\bepsilon}\}) &\propto
  \indic(\gamma^{\sf{min}}\slt\gamma\slt\gamma^{\sf{max}})
  \prod_{\bepsilon}\distBeta(\psi_{\bepsilon}\given 1, \gamma),
\end{align*}
where the products are over nodes in the aforementioned hull.  We
initialize the bounds of the slice sampler with the bounds of the top-hat
prior.

\paragraph{Selecting a Single Tree}
We have so far described a procedure for generating posterior samples from
the tree structures and associated stick-breaking processes.  If our
objective is to find a single tree, however, samples from the posterior
distribution are unsatisfying.  Following \citet{blei-etal-2010a}, we report
a best single tree structure over the data by choosing the sample from our
Markov chain that has the highest complete-data
likelihood~$p(\{x_n,\bepsilon_n\}^N_{n=1}\given\{\nu_{\bepsilon}\},
\{\psi_{\bepsilon}\}, \alpha_0,\lambda,\gamma)$.

\section{Hierarchical Clustering of Images}
\label{sec:cifar}
\begin{figure}[t!]
  \centering%
  \includegraphics[width=\textwidth]{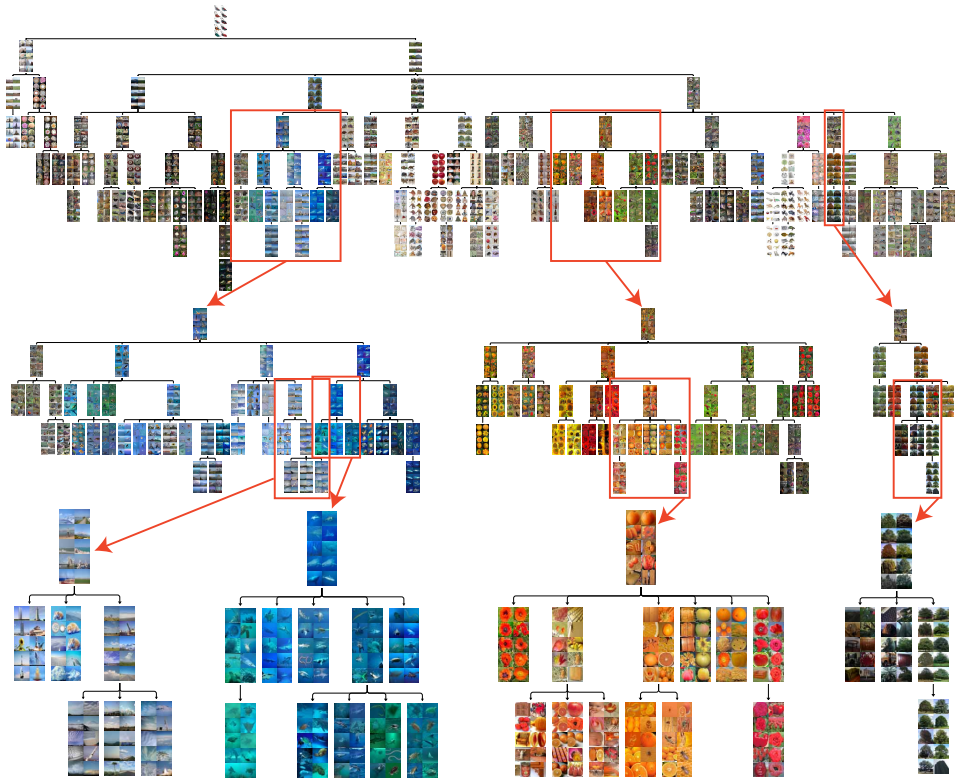}%
  \caption{These figures show a subset of the tree learned from
    the 50,000 CIFAR-100 images.  The top tree only shows nodes for
    which there were at least 250 images.  The ten shown at each node
    are those with the highest probability under the node's
    distribution.  The second row shows three expanded views of
    subtrees, with nodes that have at least 50 images.  Detailed views
    of portions of these subtrees are shown in the third row.}
  \label{fig:cifar100}
\end{figure}

We applied our model and MCMC inference to the problem of
hierarchically clustering the CIFAR-100 image data set
\footnote{\footnotesize \url{http://www.cs.utoronto.ca/~kriz/cifar.html}}.
These data are a labeled subset of the \textit{80 million tiny images}
dataset \citep{torralba-etal-2008a} with 50,000 ${32\times 32}$ color
images.  We did not use the labels in our clustering.  We modeled the
images via 256-dimensional binary features that had been extracted from
each image (i.e.,~$x_n\in\{0,1\}^{256}$) using a deep neural network that
had been trained for an image retrieval task \citep{krizhevsky-2009a}.  We
used a factored Bernoulli likelihood at each node, parameterized by a
latent 256-dimensional real vector
(i.e.,~$\theta_{\bepsilon}\in\reals^{256}$) that was transformed
component-wise via the logistic function:
\begin{align*}
  f(x_n\given \theta_{\bepsilon}) &= \prod_{d=1}^{256} \left(1+\exp\{-\theta^{(d)}_{\bepsilon}\}\right)^{-x^{(d)}_n}
  \left(1+\exp\{\theta^{(d)}_{\bepsilon}\}\right)^{1-x^{(d)}_n}.
\end{align*}
The prior over the parameters of a child node was Gaussian with its
parent's value as the mean.  The covariance of the prior ($\Lambda$ in
Section~\ref{sec:node-params}) was diagonal and inferred as part of the
Markov chain.  We placed independent~$\distUni(0.01,1)$ priors on the
elements of the diagonal.  To efficiently learn the node parameters, we
used Hamiltonian (hybrid) Monte Carlo (HMC)
\citep{duane-etal-1987a,neal-1993a}, taking 25 leapfrog HMC steps, with a
randomized step size.  We occasionally interleaved a slice sampling move
for robustness.  For the stick-breaking processes, we
used~${\alpha_0\ssim\distUni(10,50)}$, ${\lambda\ssim\distUni(0.05,0.8)}$,
and~${\gamma\ssim\distUni(1,10)}$.  Using Python on a single core of a
modern workstation each MCMC sweep of the entire model (including slice
sampled reassignment of all 50,000 images) requires approximately three
minutes.  Fig.~\ref{fig:cifar100} represents a part of the tree with the
best complete-data log likelihood after 4000 iterations.  The tree provides
a useful visualization of the data set, capturing broad variations in color
at the higher levels of the tree, with lower branches varying in texture
and shape.  A larger version of this tree is provided in the supplementary
material.

\section{Hierarchical Modeling of Document Topics}
\label{sec:nips}
We also used our approach in a bag-of-words topic model, applying it
to 1740 papers from NIPS 1--12 \footnote{\footnotesize
  \url{http://cs.nyu.edu/~roweis/data.html}}.  As in latent Dirichlet
allocation (LDA) \citep{blei-etal-2003a}, we consider a topic to be a
distribution over words and each document to be described by a
distribution over topics.  In LDA, each document has a unique topic
distribution.  In our model, however, each document lives at a node
and that \textit{node} has a unique topic distribution.  Thus multiple
documents share a distribution over topics if they inhabit the same
node.  Each node's topic distribution is from a chained
Dirichlet-multinomial as described in Section~\ref{sec:node-params}.
The topics each have symmetric Dirichlet priors over their word
distributions.  This results in a different kind of topic model than
that provided by the nested Chinese restaurant process.  In the nCRP,
each node corresponds to a topic and documents are infinitely-long
paths down the tree.  Each word is drawn from a distribution over
depths that is given by a GEM distribution.  In the nCRP, it is not
the \textit{documents} that have the hierarchy, but the
\textit{topics}.

\begin{figure}[t!]
  \centering%
  \includegraphics[width=\textwidth]{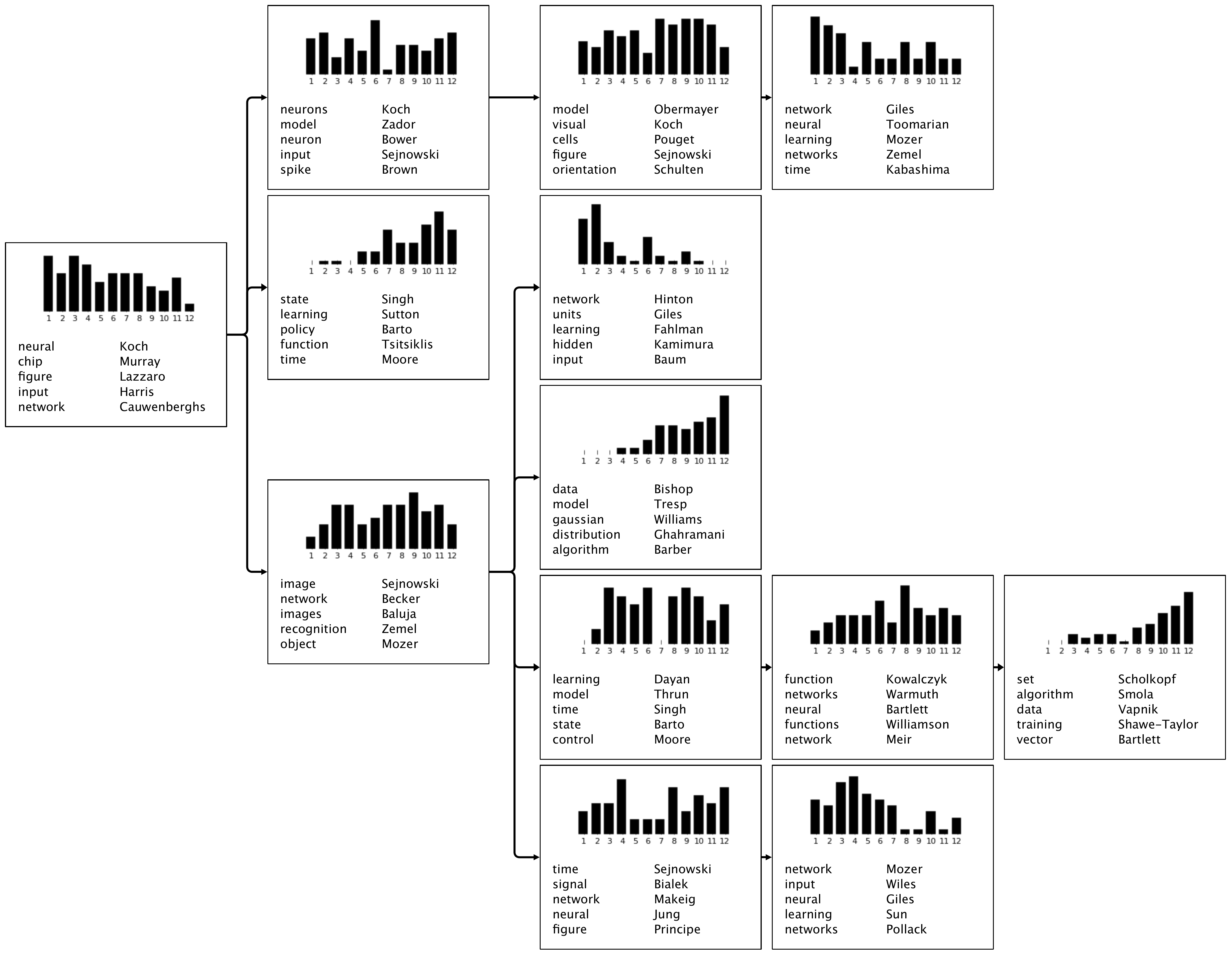}%
  \caption{A subtree of documents from NIPS 1-12, inferred
    using 20 topics.  Only nodes with at least 50 documents are shown.
    Each node shows three aggregated statistics at that node: the five most common author names,
 the five most common words and a histogram over the years of proceedings.}
  \label{fig:nips}
\end{figure}

\begin{figure}[t!]
  \centering%
  \subfloat[{Improvement versus multinomial, by number of topics}]{%
    \label{fig:nips-results:topics}%
    \includegraphics[width=0.49\textwidth]
    {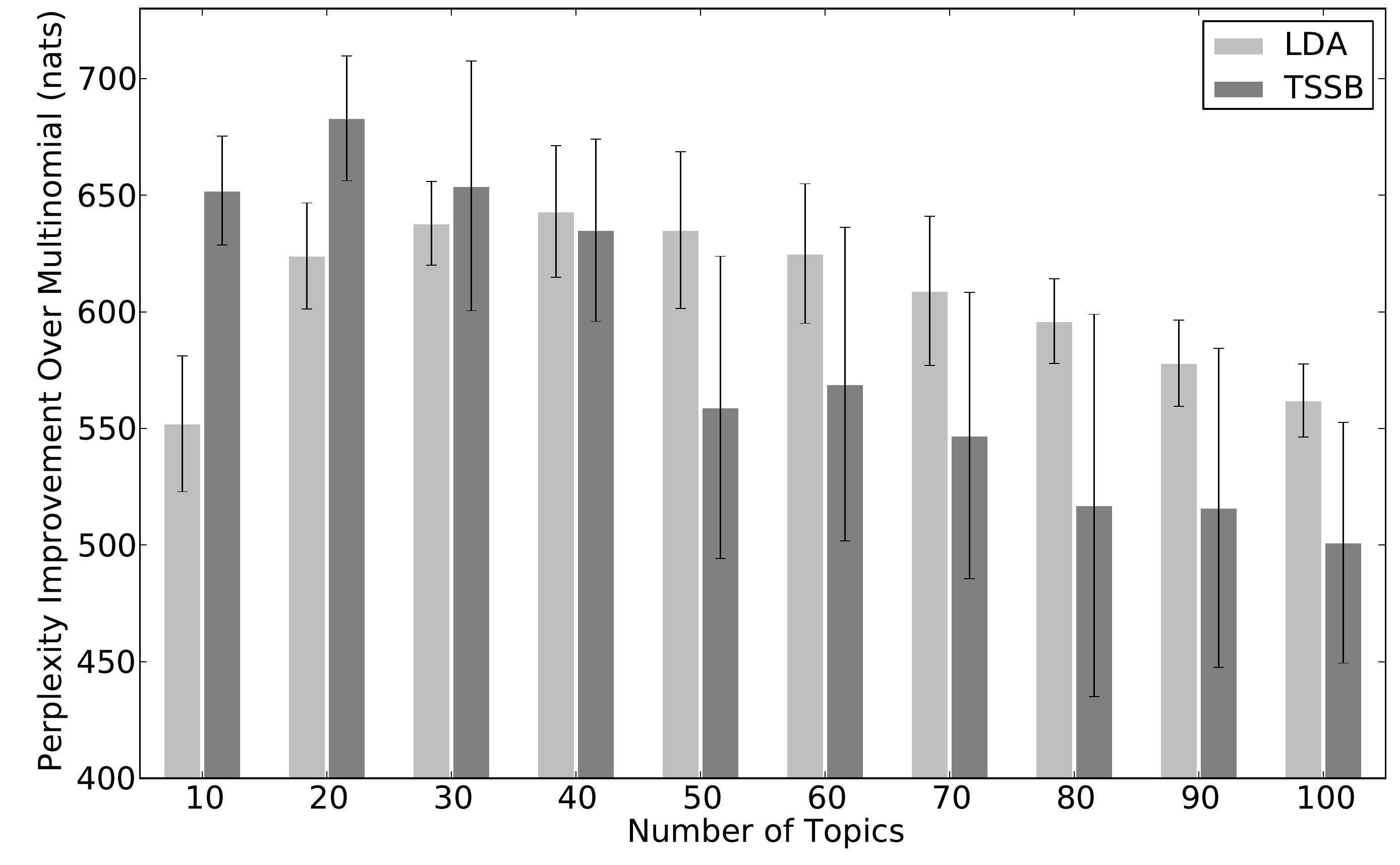}%
  }~%
  \subfloat[{Best perplexity per word, by folds}]{%
    \label{fig:nips-results:folds}%
    \includegraphics[width=0.49\textwidth]
    {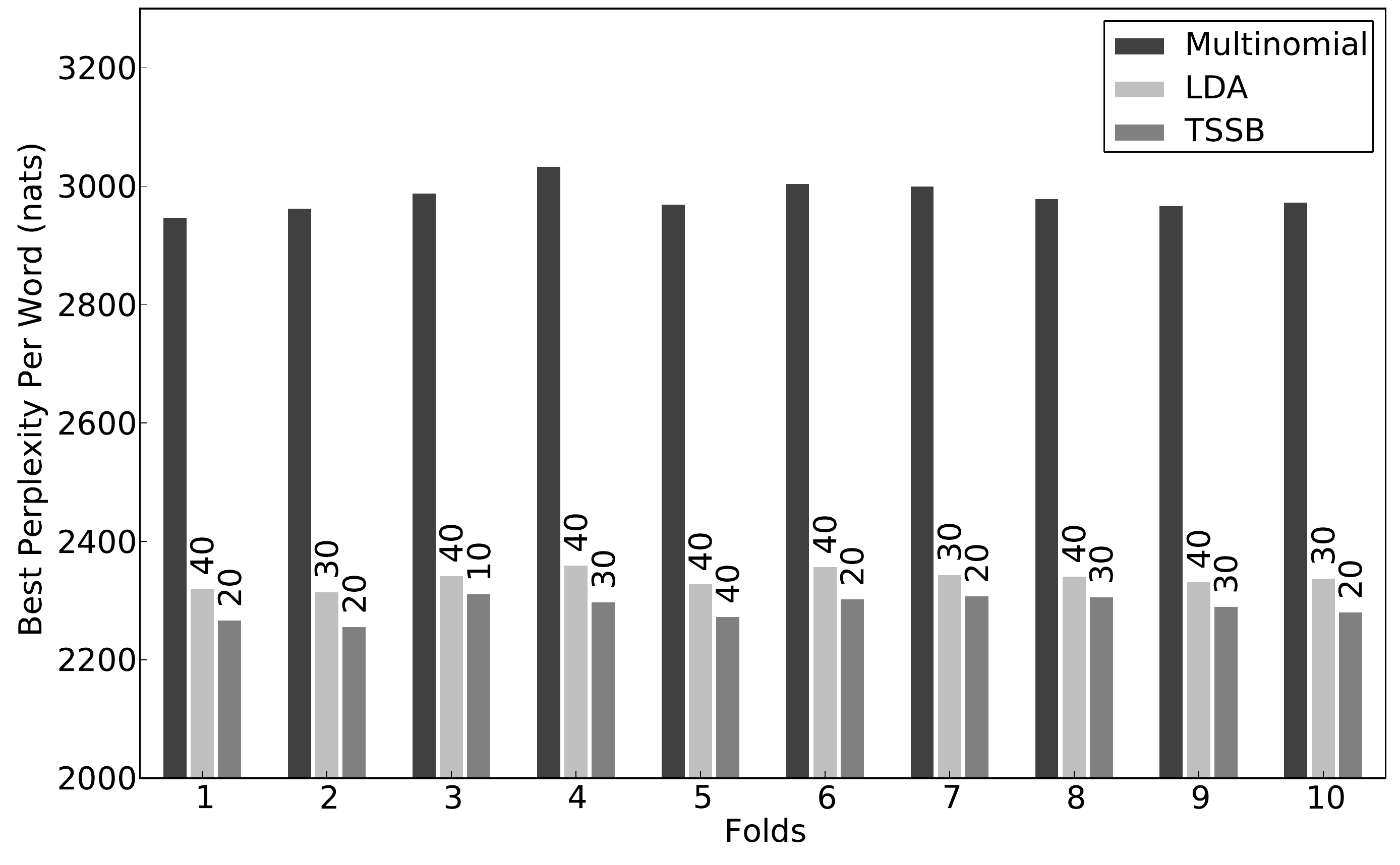}%
  }
  \caption{Results of predictive performance comparison between
    latent Dirichlet allocation (LDA) and tree-structured stick
    breaking (TSSB).  a)~Mean improvement in perplexity per word over
    Laplace-smoothed multinomial, as a function of topics (larger is
    better).  The error bars show the standard deviation of the
    improvement across the ten folds.  b)~Best predictive perplexity
    per word for each fold (smaller is better).  The numbers above the
    LDA and TSSB bars show how many topics were used to achieve this.}
  \label{fig:nips-results}%
\end{figure}

We did two kinds of analyses.  The first is a visualization as with
the image data of the previous section, using all 1740 documents.  The
subtree in Fig.~\ref{fig:nips} shows the nodes that had at least fifty
documents, along with the most common authors and words at that node.
The normalized histogram in each box shows which of the twelve years
are represented among the documents in that node.  An expanded version
of this tree is provided in the supplementary material.  Secondly, we
quantitatively assessed the predictive performance of the model.  We
created ten random partitions of the NIPS corpus into 1200 training
and 540 test documents.  We then performed inference with different
numbers of topics ($10, 20, \ldots, 100$) and evaluated the predictive
perplexity of the held-out data using an empirical likelihood estimate
taken from a mixture of multinomials (pseudo-documents of infinite
length, see, e.g.\ \citet{wallach-etal-2009b}) with 100,000 components.
As Fig.~\ref{fig:nips-results:topics} shows, our model improves in
performance over standard LDA for smaller numbers of topics.  We
believe this improvement is due to the constraints on possible topic
distributions that are imposed by the diffusion.  For larger numbers
of topics, however, it seems that these constraints become a hindrance
and the model may be allocating predictive mass to regions where it is
not warranted.  In absolute terms, more topics did not appear to
improve predictive performance for LDA or the tree-structured model.
Both models performed best with fewer than fifty topics and the best
tree model outperformed the best LDA model on all folds, as shown in
Fig.~\ref{fig:nips-results:folds}.

The MCMC inference procedure we used to train our model was as
follows: first, we ran Gibbs sampling of a standard LDA topic model
for 1000 iterations.  We then burned in the tree inference for 500
iterations with fixed word-topic associations.  We then allowed the
word-topic associations to vary and burned in for an additional 500
iterations, before drawing 5000 samples from the full posterior.  For
the comparison, we burned in LDA for 1000 iterations and then drew
5000 samples from the posterior \citep{griffiths-steyvers-2004a}.  For
both models we thinned the samples by a factor of 50.  The mixing of
the topic model seems to be somewhat sensitive to the initialization
of the~$\kappa$ parameter in the chained Dirichlet-multinomial and we
initialized this parameter to be the same as the number of topics.

\section{Discussion}
\label{sec:discussion}
We have presented a model for a distribution over random measures that also
constructs a hierarchy, with the goal of constructing a general-purpose
prior on tree-structured data.  Our approach is novel in that it combines
infinite exchangeability with a representation that allows data to live at
internal nodes on the tree, without a hazard rate process.  We have
developed a practical inference approach based on Markov chain Monte Carlo
and demonstrated it on two real-world data sets in different domains.

The imposition of structure on the parameters of an infinite mixture
model is an increasingly important topic.  In this light, our notion
of evolutionary diffusion down a tree sits within the larger class
of models that construct dependencies between distributions on random
measures~\citep{maceachern-1999a,maceachern-etal-2001a,teh-etal-2006a}.

\subsection*{Acknowledgements}
The authors wish to thank Alex Krizhevsky for providing the CIFAR-100
binary features; Iain Murray and Yee Whye Teh for valuable advice
regarding MCMC inference; Kurt Miller, Hanna Wallach and Sinead
Williamson for helpful discussions.  RPA is a junior fellow of the
Canadian Institute for Advanced Research.

\bibliographystyle{abbrvnat}
\bibliography{draft}

\begin{thebibliography}{29}
\providecommand{\natexlab}[1]{#1}
\providecommand{\url}[1]{\texttt{#1}}
\expandafter\ifx\csname urlstyle\endcsname\relax
  \providecommand{\doi}[1]{doi: #1}\else
  \providecommand{\doi}{doi: \begingroup \urlstyle{rm}\Url}\fi

\bibitem[Bart et~al.(2008)Bart, Porteous, Perona, and Welling]{bart-etal-2008a}
E.~Bart, I.~Porteous, P.~Perona, and M.~Welling.
\newblock Unsupervised learning of visual taxonomies.
\newblock In \emph{{IEEE} Conference on Computer Vision and Pattern
  Recognition}, 2008.

\bibitem[Blei et~al.(2003)Blei, Ng, and Jordan]{blei-etal-2003a}
D.~M. Blei, A.~Y. Ng, and M.~I. Jordan.
\newblock Latent {D}irichlet allocation.
\newblock \emph{Journal of Machine Learning Research}, 3:\penalty0 993--1022,
  2003.

\bibitem[Blei et~al.(2010)Blei, Griffiths, and Jordan]{blei-etal-2010a}
D.~M. Blei, T.~L. Griffiths, and M.~I. Jordan.
\newblock The nested {C}hinese restaurant process and {B}ayesian nonparametric
  inference of topic hierarchies.
\newblock \emph{Journal of the ACM}, 57\penalty0 (2):\penalty0 1--30, 2010.

\bibitem[{Daum\'e III}(2009)]{daume-2009a}
H.~{Daum\'e III}.
\newblock {B}ayesian multitask learning with latent hierarchies.
\newblock In \emph{Proceedings of the 25th Conference on Uncertainty in
  Artificial Intelligence}, 2009.

\bibitem[Duane et~al.(1987)Duane, Kennedy, Pendleton, and
  Roweth]{duane-etal-1987a}
S.~Duane, A.~D. Kennedy, B.~J. Pendleton, and D.~Roweth.
\newblock Hybrid {M}onte {C}arlo.
\newblock \emph{Physics Letters B}, 195\penalty0 (2):\penalty0 216--222, 1987.

\bibitem[Griffiths and Steyvers(2004)]{griffiths-steyvers-2004a}
T.~L. Griffiths and M.~Steyvers.
\newblock Finding scientific topics.
\newblock \emph{Proceedings of the National Academy of Sciences of the United
  States of America}, 101\penalty0 (Suppl. 1):\penalty0 5228--5235, 2004.

\bibitem[Heller and Ghahramani(2005)]{heller-ghahramani-2005a}
K.~A. Heller and Z.~Ghahramani.
\newblock {B}ayesian hierarchical clustering.
\newblock In \emph{Proceedings of the 22nd International Conference on Machine
  Learning}, 2005.

\bibitem[Ishwaran and James(2001)]{ishwaran-james-2001a}
H.~Ishwaran and L.~F. James.
\newblock {G}ibbs sampling methods for stick-breaking priors.
\newblock \emph{Journal of the American Statistical Association}, 96\penalty0
  (453):\penalty0 161--173, March 2001.

\bibitem[Kemp et~al.(2004)Kemp, Griffiths, Stromsten, and
  Tenenbaum]{kemp-etal-2004a}
C.~Kemp, T.~L. Griffiths, S.~Stromsten, and J.~B. Tenenbaum.
\newblock Semi-supervised learning with trees.
\newblock In \emph{Advances in Neural Information Processing Systems 16}. 2004.

\bibitem[Krizhevsky(2009)]{krizhevsky-2009a}
A.~Krizhevsky.
\newblock Learning multiple layers of features from tiny images.
\newblock Technical report, Department of Computer Science, University of
  Toronto, 2009.

\bibitem[MacEachern(1999)]{maceachern-1999a}
S.~N. MacEachern.
\newblock Dependent nonparametric processes.
\newblock In \emph{Proceedings of the Section on {B}ayesian Statistical
  Science}, 1999.

\bibitem[MacEachern et~al.(2001)MacEachern, Kottas, and
  Gelfand]{maceachern-etal-2001a}
S.~N. MacEachern, A.~Kottas, and A.~E. Gelfand.
\newblock Spatial nonparametric {B}ayesian models.
\newblock Technical Report 01-10, Institute of Statistics and Decision
  Sciences, Duke University, 2001.

\bibitem[Mauldin et~al.(1992)Mauldin, Sudderth, and
  Williams]{mauldin-etal-1992a}
R.~D. Mauldin, W.~D. Sudderth, and S.~C. Williams.
\newblock {P\'{o}lya} trees and random distributions.
\newblock \emph{The Annals of Statistics}, 20\penalty0 (3):\penalty0
  1203--1221, September 1992.

\bibitem[Meeds et~al.(2008)Meeds, Ross, Zemel, and Roweis]{meeds-etal-2008a}
E.~Meeds, D.~A. Ross, R.~S. Zemel, and S.~T. Roweis.
\newblock Learning stick-figure models using nonparametric {B}ayesian priors
  over trees.
\newblock In \emph{{IEEE} Conference on Computer Vision and Pattern
  Recognition}, 2008.

\bibitem[Neal(1993)]{neal-1993a}
R.~M. Neal.
\newblock Probabilistic inference using {M}arkov chain {M}onte {C}arlo methods.
\newblock Technical Report CRG-TR-93-1, Department of Computer Science,
  University of Toronto, 1993.

\bibitem[Neal(2003{\natexlab{a}})]{neal-2003a}
R.~M. Neal.
\newblock Density modeling and clustering using {D}irichlet diffusion trees.
\newblock In \emph{Bayesian Statistics 7}, pages 619--629, 2003{\natexlab{a}}.

\bibitem[Neal(2003{\natexlab{b}})]{neal-2003b}
R.~M. Neal.
\newblock Slice sampling (with discussion).
\newblock \emph{The Annals of Statistics}, 31\penalty0 (3):\penalty0 705--767,
  2003{\natexlab{b}}.

\bibitem[Papaspiliopoulos and Roberts(2008)]{papaspiliopoulos-roberts-2008a}
O.~Papaspiliopoulos and G.~O. Roberts.
\newblock Retrospective {M}arkov chain {M}onte {C}arlo methods for {D}irichlet
  process hierarchical models.
\newblock \emph{Biometrika}, 95\penalty0 (1):\penalty0 169--186, 2008.

\bibitem[Pitman(1996)]{pitman-1996a}
J.~Pitman.
\newblock Random discrete distributions invariant under size-biased
  permutation.
\newblock \emph{Advances in Applied Probability}, 28\penalty0 (2):\penalty0
  525--539, 1996.

\bibitem[Pitman(2002)]{pitman-2002a}
J.~Pitman.
\newblock Poisson--{D}irichlet and {GEM} invariant distributions for
  split-and-merge transformation of an interval partition.
\newblock \emph{Combinatorics, Probability and Computing}, 11:\penalty0
  501--514, 2002.

\bibitem[Pitman and Yor(1997)]{pitman-yor-1997a}
J.~Pitman and M.~Yor.
\newblock The two-parameter {P}oisson--{D}irichlet distribution derived from a
  stable subordinator.
\newblock \emph{The Annals of Probability}, 25\penalty0 (2):\penalty0 855--900,
  1997.

\bibitem[Roy et~al.(2007)Roy, Kemp, Mansinghka, and Tenenbaum]{roy-etal-2007a}
D.~M. Roy, C.~Kemp, V.~K. Mansinghka, and J.~B. Tenenbaum.
\newblock Learning annotated hierarchies from relational data.
\newblock In \emph{Advances in Neural Information Processing Systems 19}, 2007.

\bibitem[Sethuraman(1994)]{sethuraman-1994a}
J.~Sethuraman.
\newblock A constructive definition of {D}irichlet priors.
\newblock \emph{Statistica Sinica}, 4:\penalty0 639--650, 1994.

\bibitem[Teh et~al.(2006)Teh, Jordan, Beal, and Blei]{teh-etal-2006a}
Y.~W. Teh, M.~I. Jordan, M.~J. Beal, and D.~M. Blei.
\newblock Hierarchical {D}irichlet processes.
\newblock \emph{Journal of the American Statistical Association}, 101\penalty0
  (476):\penalty0 1566--1581, 2006.

\bibitem[Teh et~al.(2007)Teh, {Daum\'e III}, and Roy]{teh-etal-2007a}
Y.~W. Teh, H.~{Daum\'e III}, and D.~Roy.
\newblock {B}ayesian agglomerative clustering with coalescents.
\newblock In \emph{Advances in Neural Information Processing Systems 20}, 2007.

\bibitem[Torralba et~al.(2008)Torralba, Fergus, and
  Freeman]{torralba-etal-2008a}
A.~Torralba, R.~Fergus, and W.~T. Freeman.
\newblock 80 million tiny images: A large data set for nonparametric object and
  scene recognition.
\newblock \emph{{IEEE} Transactions on Pattern Analysis and Machine
  Intelligence}, 30\penalty0 (11):\penalty0 1958--1970, 2008.

\bibitem[Walker(2007)]{walker-2007a}
S.~G. Walker.
\newblock Sampling the {D}irichlet mixture model with slices.
\newblock \emph{Communications in Statistics}, 36:\penalty0 45--54, 2007.

\bibitem[Wallach et~al.(2009)Wallach, Murray, Salakhutdinov, and
  Mimno]{wallach-etal-2009b}
H.~M. Wallach, I.~Murray, R.~Salakhutdinov, and D.~Mimno.
\newblock Evaluation methods for topic models.
\newblock In \emph{Proceedings of the 26th International Conference on Machine
  Learning}, 2009.

\bibitem[Williams(2000)]{williams-2000a}
C.~K.~I. Williams.
\newblock A {MCMC} approach to hierarchical mixture modelling.
\newblock In \emph{Advances in Neural Information Processing Systems 12}, pages
  680--686. 2000.

\end{thebibliography}

\end{document}